\documentclass{cernrep}
\newcommand{\beq}{\begin{equation}}
\newcommand{\eeq}{\end{equation}}
\newcommand{\beqa}{\begin{eqnarray}}
\newcommand{\eeqa}{\end{eqnarray}}
\newcommand{\no}{\nonumber}
\newcommand{\slam}{\slash \hspace{-0.25cm}}
\newcommand{\varep}{\varepsilon}
\newcommand{\mtilde}{\tilde{m}}

\newcommand{\tr}{{\mathrm{Tr}}}
\begin{document}
\title{Introduction to Supersymmetry}
\author{Y.~Shadmi}
\institute{Technion---Israel Institute of Technology, Haifa 32100, Israel}
\maketitle

\begin{abstract}
These lectures are a brief introduction to supersymmetry.
\end{abstract}

\section{Introduction}\label{intro}
You have probably heard in the past about the motivation for supersymmetry.
Through these lectures, this will (hopefully) become
clear and more concrete.
But it's important to state at the outset:
There is no experimental evidence for supersymmetry.
The amount of effort that has been invested in supersymmetry, in both
theory and experiment,
may thus be somewhat surprising.
In this respect, supersymmetry is no different from any other ``new physics'' scenario.
There is no experimental evidence for {\sl any} underlying theory of
electroweak symmetry breaking, which would give rise to the 
(fundamental scalar) Higgs mechanism as an effective description.
There is of course experimental evidence for physics beyond the standard model 
(SM):
 dark matter, 
the baryon asymmetry---CP violation.

Why so much effort on supersymmetry?
It is a very beautiful and exciting idea
because
it's conceptually different from anything we know
in Nature.
It's a symmetry that relates particles of 
{\sl different spins}---bosons and fermions.
In fact, we are now in a very special era from the point of view of spin:
The Higgs was discovered.
As far as we know, it is a fundamental particle. So for the first time, 
we have a spin-0 fundamental particle.
It would be satisfying to have some unified understanding
of the spins we observe.
Supersymmetry would be a step in this direction.
Given the SM fermions, it predicts 
spin-0 particles.
Beyond this purely theoretical motivation, the fact that the Higgs is
a scalar poses a more concrete (yet purely theoretical)
puzzle.
Scalar fields (unlike vector bosons or fermions) have quadratic divergences.
This leads to the fine-tuning problem, as we will review in the next Section. 
Supersymmetry removes these divergences. 
We will see that in some sense {\bf Supersymmetry makes a
scalar behave like a fermion.}

Supersymmetry is not a specific model.
Rather, there is a wide variety of supersymmetric extensions of the SM.
These involve
different superpartner spectra, and therefore different experimental signatures.
In thinking about these, we have developed  a whole toolbox for collider 
searches, including different triggers
and analyses. 
In particular, supersymmetry supplies many concrete examples with
new scalars (same charges as SM fermions),
new fermions (same charges as SM gauge bosons),
potentially leading to  
missing energy,
displaced vertices,
long lived charged particles,
or disappearing tracks, to name just some of the possible signatures..
For discovery, spin is a secondary consideration.
So even if we are misguided in thinking about supersymmetry, 
and Nature is not supersymmetric, 
the work invested in supersymmetry searches may help us discover something else.

%
\subsection{Plan}

In Section~\ref{basics} of these lectures we will see the basics of 
supersymmetry
through a few simple toy models. 
These toy models can be thought of as ``modules'' for building the
supersymmetrized standard model.
In the process, we will try to de-mystify supersymmetry and understand the following questions about it:
\begin{itemize}
\item In what sense is it a space-time symmetry (extending translations, rotations and boosts)?
\item Why does it remove UV divergences (thus solving the fine-tuning, or Naturalness problem)?
\item Why do we care about it even though it's clearly broken?
\item Why is the gravitino relevant for LHC experiments?
\end{itemize}
%
%
In Section~\ref{ssm} we will describe the Minimal Supersymmetric Standard
Model (MSSM).
Here we will put to use what we learned in Section~\ref{basics}.
We will discuss:
\begin{itemize}
\item Motivation (now that you can appreciate it)
\item The field content
\item The interactions: we will see that there is (almost) no freedom in these
\item The supersymmetry-breaking terms: this is where we have freedom, and these
determine experimental signatures. 
\item EWSB and the Higgs mass
\item Spectra (the general structure of superpartner masses)
\end{itemize}
%
%
We will conclude in Section~\ref{lhc} with general considerations
for LHC searches. 

The aim of these lectures is to provide a conceptual understanding of 
supersymmetry and the supersymmetrized standard model.
Therefore, we will start with a pretty technical review of what symmetries
are and their derivation from Lagrangians. This is necessary so that
you get a clear idea of what supersymmetry is. As we go on however, we will
take a more qualitative approach. For more details, I refer you to the 
many excellent books and reviews of the subject, including, 
for example~\cite{Baer:2006rs,Dine:1998ia,Martin:1997ns}.
I cannot do justice to the vast literature on the subject in these 
short lectures. For original references, I refer you again to the books
and reviews above.

\section{Supersymmetry basics}\label{basics}
\subsection{Spacetime symmetry}
The symmetry we are most familiar with is
Poincare symmetry. It contains
\begin{itemize}
\item Translations: $x^\mu\to x^\mu+a^\mu$~~~(generators: $P^\mu$) 
\item
Lorentz transformations: $x^\mu\to x^\mu +w^\mu_\nu x^\nu$, 
where $w^{\mu\nu}$ is antisymmetric~~~(generators: $J^{\mu\nu}$)
\end{itemize}
Throughout we will only consider global, infinitesimal transformations, so
$a^\mu$ and $w^{\mu\nu}$ are small, coordinate-independent numbers.

These transformations contain rotations. For a rotation around 
the axis $x^k$ (with angle $\theta^k$):
$w^{ij}=\epsilon^{ijk} \theta^k$.
Thus for example for rotations around $z$:
\beq
x^0\to x^0\,;~~~
x^1\to x^1-\theta x^2 \,;~~~
x^2\to x^2+\theta x^1 \,;~~~
x^3\to x^3\no
\eeq

The Lorentz transformations also contain
boosts. For a boost
along the axis $x^k$ (with speed $\beta^k$): $-w^{0k}=w^{k0}=\beta^k$.
Thus for example for a boost along $z$,
\beq
x^0\to x^0+\beta x^3\,;~~~
\,;~~~x^1\to x^1\,;~~~
x^2\to x^2\,;~~~
x^3\to x^3 +\beta x^0\no\,.
\eeq

The Lorentz algebra is:
\beq
[P^\mu,P^\nu]= 0\no
\eeq
\beq
  [P^\mu,J^{\rho\sigma} ]=   0
\eeq
\beq
[J^{\mu\nu},J^{\rho\sigma} ]=
i  (g^{\nu\rho} J^{\mu\sigma}
-g^{\mu\rho} J^{\nu\sigma}-g^{\nu\sigma} J^{\mu\rho}+g^{\mu\sigma} J^{\nu\rho})\,.
\no
\eeq
Here $P^\mu$ are the momenta---the generators of translations, 
$J^{\mu\nu}$ contain  the angular momenta, which generate
rotations (with $\mu,\nu=1,2,3$) and the generators of boosts  (with $\mu=0,\nu=1,2,3$).

Let's recall where all this is coming from. To do that,
let's ``discover'' all the above in a simple field theory,
namely a field theory of a single
free complex scalar field.
The Lagrangian is,
\beq
{\cal L}= 
\partial^\mu\phi^* \,\partial_\mu\phi -
m^2 \left\vert \phi \right\vert^2 \ .
\eeq  

What's a symmetry? It's a transformation of the fields which leaves the Equations 
of Motion (EOMs) invariant.
The EOMs follow from the action, so this tells us that the action is invariant under 
the symmetry transformation.
Since the action is the integral of the Lagrangian, it follows that the
Lagrangian can change by a total derivative,
\beq
{\cal L}\to {\cal L} + \alpha \partial_\mu{\cal J}^\mu\,,
\eeq
where $\alpha$ is the (small) parameter of the transformation.

What's the symmetry of our toy theory?
First, there's a U(1) symmetry:
\beq
\phi(x)\to e^{i\alpha} \phi(x)\,.
\eeq
Under this transformation, 
${\cal L}$ is invariant.
This is an example of
an {\sl internal symmetry}, that is, a symmetry which is {\sl not} a space-time symmetry 
(it does not do anything to the coordinates).

But our toy theory also has {\sl spacetime symmetries}:\\
{\bf Translations:} 
\beqa
x^\mu&\to& x^\mu+a^\mu\\
\phi(x)\to \phi(x-a)&=&\phi(x) -a^\mu \partial_\mu \phi(x)
\eeqa
or
\beq
\delta_a\phi(x)=a^\mu \partial_\mu \phi(x)
\eeq
and
{\bf Lorentz transformations:}
\beqa
x^\mu&\to& x^\mu +w^{\mu\nu} x_\nu\\
\phi(x^\mu)&\to& \phi(x^\mu-w^{\mu\nu}x_\nu) 
\eeqa
so
\beq
\delta_w\phi(x)=w^{\mu\nu} x_\mu\partial_\nu \phi(x)
=\frac12\,w^{\mu\nu} (x_\mu\partial_\nu- x_\nu\partial_\mu)\,\phi(x)
\eeq
The Lagrangian only changes by a total derivative under these, so the action is invariant.

Now let's see how the algebra arises. Consider performing two translations.
First we do a translation with $x^\mu\to x^\mu+a^\mu$. 
Then we do a   translation with $x^\mu\to x^\mu+b^\mu$.
Alternatively, we could first perform the translation with $b^\mu$ and then the one with $a^\mu$.
Obviously, this should not make any difference.
Mathematically, this translates to the fact that the commutator of two translations vanishes. 
Indeed,
\beq
[\delta_a,\delta_b]\phi\equiv \delta_a(\delta_b\phi)-\delta_b(\delta_a\phi)=0\,.
\eeq
With rotations and boosts, the order does matter. 
Consider the commutator of two Lorentz transformations with parameters 
$w^{\mu\nu}$ and $\lambda^{\rho\sigma}$:
\beq
[\delta_{w^{\mu\nu}},\delta_{\lambda^{\rho\sigma}}]\phi= i 
w_{\mu\nu}\lambda_{\rho\sigma}\, \cdot i\,
\left\{g^{\nu\rho} (x_\mu\partial_\sigma- x_\mu\partial_\sigma) +~{\rm permutations}
    \right\}
\eeq
So we derived the algebra of spacetime symmetry  transformations (=the Poincare algebra) 
in this toy example.
Now let's do the same in a supersymmetric theory.

\subsection{A simple supersymmetric field theory}
Our example will be a
free theory with one massive (Dirac) fermion  of mass $m$, which we will denote by $\psi(x)$,
and two complex scalars of mass $m$ which we will denote by $\phi_+(x)$ and  $\phi_-(x)$. 
The Lagrangian is,
\beq
{\cal L}= 
\partial^\mu\phi_+^* \,\partial_\mu\phi_+ -
m^2 \left\vert \phi_+ \right\vert^2 
+\partial^\mu\phi_-^* \,\partial_\mu\phi_- -
m^2 \left\vert \phi_- \right\vert^2 
+ \bar\psi(i\slam\partial-m)\psi 
\eeq  
The labels $+$, $-$ are just names, we'll see the reason for this choice soon.
This isn't the most  minimal supersymmetric 4d field theory.
``Half of it'' is: a 2-component (Weyl) fermion plus one complex scalar.
But Dirac spinors are more familiar so we start with this example.

Just as in the previous example, this theory has
spacetime symmetry, including translations,
rotations, and boosts.
The only difference is that
$\psi(x)$ itself is a spinor, so it transforms nontrivially,
\beq
\psi(x)\to \psi^\prime(x^\prime)\,.
\eeq
Actually, the L-handed and R-handed parts of the spinor transform differently under Lorentz.
Write
\beqa
\begin{pmatrix}
\psi_L\\
\psi_R 
\end{pmatrix}
\eeqa
where $\psi_L$ and $\psi_R$ are 2-component spinors.
Then under Lorentz transformations,
\beqa
\psi_L&\to& \psi^\prime_L= (1-i\theta^i \frac{\sigma^i}2 
- \beta^i \frac{\sigma^i}2)\\
\psi_R&\to& \psi^\prime_R= (1-i\theta^i \frac{\sigma^i}2 
+ \beta^i \frac{\sigma^i}2)
\eeqa
so it will be useful to write everything in terms of 2-component spinors.

Recall that 
we can write any R-handed spinor in terms of a L-handed one:
\beq\label{lr}
\psi_R= -\varep \chi_L^*
\eeq
where
\beqa
\varep\equiv -i\sigma^2 =
\begin{pmatrix}
0 &-1\\
1& 0
\end{pmatrix}
\eeqa

{\bf Exercise:} prove eq.~(\ref{lr})

We can then write our Dirac spinor in terms of two  {\bf L-handed spinors}
$\psi_+$ and $\psi_-$:
$\psi_L=\psi_+$, $\psi_R=-\varep \psi_-^*$ so that,
\beqa
 \psi=
\begin{pmatrix}
\psi_-\\
-\varep \psi_+^* 
\end{pmatrix}
\eeqa

Let's write the Lagrangian in terms of these 2-component spinors,
\beqa\label{toy}
{\cal L}&=& 
\partial^\mu\phi_+^* \,\partial_\mu\phi_+ 
+ \psi_+^\dagger i\bar\sigma^\mu \partial_\mu \psi_+\no\\
&+&\partial^\mu\phi_-^* \,\partial_\mu\phi_- +
\psi_-^\dagger i\bar\sigma^\mu \partial_\mu \psi_-\\
&-&m^2 \left\vert \phi_- \right\vert^2 
-m^2 \left\vert \phi_+ \right\vert^2 
-m ( \psi_+^T\varep \psi_- + {\rm hc})\no
\eeqa  

{\bf Exercise:}
Derive this. Show also that
$\psi_+^T\varep \psi_- = \psi_-^T\varep \psi_+$, 
where $\psi_\pm$ are any 2-component spinors.

All we've done so far is to re-discover spacetime symmetry in this simple field theory.
Now comes the big question:
{\bf Can this spacetime symmetry be extended? }

The answer is YES: there's more symmetry hiding in our theory!
Take a constant (anti-commuting)  2-component L-spinor $\xi$.
Consider the following transformations,
\beqa
\delta_\xi \phi_+ &=&\sqrt2\, \xi^T \varep \psi_+\nonumber\\
\delta_\xi \psi_+ &=&\sqrt2\, i\sigma^\mu  \varep \xi^*\partial_\mu\phi_+
\eeqa
and similarly for $+\to -$.

{\bf Exercise:}
Check that this is  a symmetry of our theory:
1. Show that the massless part of the Lagrangian is invariant.
2. Show that the rest of the Lagrangian is invariant too {\sl if}
the masses of the fermion and scalars are the same.
Here you will have to use the EOMs.

We see that
{\sl the symmetry transformations take a boson into a fermion and vice versa.}
THIS IS SUPERSYMMETRY.

As an (important) aside, we
note that the symmetry {\sl separately} relates $\phi_--\psi_-$ and $\phi_+-\chi_+$.
Thus,  if $m=0$, the two  halves of the theory decouple, and each one is symmetric separately.
Therefore, as mentioned above, this theory is not the most minimal supersymmetric theory,
but half of it is. 
This is very handy if we're to implement supersymmetry in the SM, 
because the SM is a {\sl chiral
theory}.

Is the symmetry we found  indeed an extension of Poincare?
It's surely a spacetime symmetry since it takes a fermion into a boson
(the transformation parameters carry spinor indices).
Furthermore, let's consider the algebra.
Take the commutator of two new transformations with parameters $\xi$,
$\eta$:
\beq\label{susyalg}
[\delta_\xi,\delta_\eta]\phi_L =  a^\mu \partial_\mu\phi_L
~~~~~~~~~
{\mathrm with}
~~~~
a^\mu= 2i \left(\xi^\dagger \bar\sigma^\mu \eta -
\eta^\dagger \bar\sigma^\mu \xi
\right)
\eeq
This is a translation!
We see that the commutator of two new transformations gives a translation.
So indeed, the new symmetry is an extension of the ``usual'' spacetime symmetry.

{\bf Exercise:} 
Check eq.~(\ref{susyalg}). You will have to use the EOMs.

Let's summarize:
Our simple theory is supersymmetric. 
We have an extension of spacetime symmetry
that involves anti-commuting generators. 
The supersymmetry transformations relate bosons and fermions.

There are a couple of features of this simple example that are worth stressing 
because they hold quite generally: (i)~If the bosons and fermions had different masses, 
we would not have this symmetry. That is,  
the theory would not be  supersymmetric.
(ii) Let's count the physical degrees of freedom (dof's): on-shell we have
$2+2=4$ fermions, 
$2+2=4$ bosons. Thus we have equal numbers of fermionic and bosonic dof's.
(Off shell, the bosons are the same, but the fermions have $2\times 4$.)

\subsection{The vacuum energy}
Recall that global symmetries lead to Noether currents.
For each global symmetry there is a
current $j^\mu$, with $\partial_\mu j^\mu=0$,
 so that there is a
 conserved charge: 
\beq
Q = \int d^3x\, j^0(x)~~~~{\rm with}~~~
\frac{d}{dt}\, Q=0 \,.
\eeq
For translations in time, 
the conserved charge is the Hamiltonian $H$.

Thus, what we found above in eqn.~(\ref{susyalg}) means that the anti-commutator of 
two supersymmetry transformations gives the Hamiltonian.
Schematically,
\beq
\left\{
{\mathrm{SUSY, SUSY}} \right\}\propto H\,,
\eeq
where SUSY stands for the generator of a supersymmetry transformation.

Now consider the vacuum expectation value (VEV) of this last relation,
\beq\label{hamilt}
\langle0\vert\left\{
{\mathrm{SUSY, SUSY}}
\right\}
 \vert0\rangle\propto \langle0\vert H\vert0\rangle\,.
\eeq
If supersymmetry is unbroken, the ground state is supersymmetric. Therefore it is annihilated by the SUSY
generator, 
\beq
{\mathrm{SUSY}} \vert0\rangle=0 \,.
\eeq
Using eqn~(\ref{hamilt}) we find
\beq\label{vacen}
\langle0\vert H\vert0\rangle=0\,.
\eeq
In a supersymmetric theory, the ground state energy is  zero.

As you probably heard many times (and as we will review soon), 
one of the chief motivations for supersymmetry is the fine-tuning problem, that is,
the fact that supersymmetry removes the quadratic divergence in the Higgs mass.
Here you can already see the power of supersymmetry in removing UV divergences.
The vacuum energy usually diverges. 
(This should remind you of your first quantum mechanics class, where you saw that there's
an infinite constant in the energy 
of the harmonic oscillator. In QM, we just set this infinite constant
to zero, by choosing the zero of the energy.)
This is in fact the worst divergence we encounter
in field theory, a  quartic divergence.
Now we see that supersymmetry completely removes this divergence:
in a supersymmetric theory, the ground state energy is zero.
This gives us hope that supersymmetry can help with other UV divergences.

%

The  next worst divergence you can have in field theory is a quadratic
divergence. Where does it show up? In the mass-squared of scalar fields: 
\beq
\delta m^2 \propto \Lambda^2
\eeq
where $\Lambda$ is the cutoff.
This is why we are worried about fine tuning in the Higgs mass.
You could ask yourself why no one ever worries about the electron mass. 
It too is  much smaller than the Planck scale.
The reason this is not a problem, is that
fermion masses have no quadratic divergences,
only logarithmic divergences.
This is a very important result so we will see it in three ways.\\
(1) Consider a fermion Lagrangian with a mass term $m_0$,
\beq
{\cal L}= \bar\psi(i\slam\partial-m_0)\psi ~
= ~\bar\psi(i\slam\partial)\psi -m_0 (\psi_L^\dagger \psi_R 
+ \psi_R^\dagger \psi_L)
\eeq
Note that the mass term is the only term that couples $\psi_L$ and $\psi_R$.
So if $m_0=0$, $\psi_L$, $\psi_R$ don't talk to each other.
A mass term (L-R coupling) is never generated.
Therefore, even if we include quantum corrections,
\beq
\delta m \propto m_0\,,
\eeq
(Here $m$ is the full, physical mass including quantum corrections.)
We see that
with $m_0=0$ we have two different species: $\psi_L$---call it,say, a ``blue'' fermion,
and $\psi_R$, a ``red fermion'', and they don't interact at all.

(2)
Consider $m_0\neq0$.
Take a L-fermion (spin along $\hat p$). This is our  ``blue fermion''.
We can run very fast alongside. If our speed is greater than the fermion's speed, 
$\hat p \to -\hat p$, but the spin stays the same.
Thus the fermion helicity (which is the projection of the spin along the direction of motion)
changes. L becomes R.
The ``blue'' fermion turns into a ``red'' fermion.
(We see that helicity is not a good quantum number for a massive fermion.)
But if $m_0=0$,  the ``blue'' fermion is massless. It travels at the speed of 
light---we can never run fast enough. The blue fermion does not change into a red fermion.
Thus, L and R are distinct in this case, and
the blue fermion and the red fermion are decoupled.
%
We thus learn that any correction to the bare mass $m_0$ must be proportional to $m_0$,
\beq
\delta m \equiv m-m_0 \propto m_0 
\eeq
How can the UV cutoff $\Lambda$ enter?
On dimensional grounds,
\beq\label{mpropm0}
\delta m \propto m_0 \, \log\frac{m_0}{\Lambda}\,,
\eeq
so 
\beq
\delta m = 0\cdot \Lambda + \#\, m_0 \, \log\frac{m_0}{\Lambda}\,.
\eeq
Indeed, there is no quadratic divergence in the fermion mass.
The worst divergence that can appear is a logarithmic divergence.
This is why no one ever worries about fine-tuning in the electron mass.

(3)
Again, the question we're asking is:
why is there no quadratic divergence in the fermion mass?
We'll now see this using a global symmetry---the
chiral symmetry.
Let's consider the fermion Lagrangian again,
\beq
{\cal L}=
 \bar\psi(i\slam\partial)\psi -m_0 (\psi_L^\dagger \psi_R + \psi_R^\dagger \psi_L)
\eeq
if $m_0=0$, we have two independent U(1) symmetries, U(1)$_L\times$U(1)$_R$.
This symmetry forbids the mass term. We again conclude~(\ref{mpropm0}).


We saw that 
supersymmetry implies that the
boson mass equals the fermion mass.
We also saw that
chiral symmetry implies that there is no quadratic divergence in the fermion mass.
Putting these together we conclude that
{\bf in a supersymmetric theory: there is no quadratic divergence  in the boson mass.}
This is how supersymmetry solves the fine-tuning problem.

But there's more that we can learn just based on dimensional analysis.
We know there is no supersymmetry in Nature. We know for example that
there is no spin-0 particle whose mass  equals the electron mass.
So why should we care about supersymmetry?
The reason is that supersymmetry is so powerful that even when it's broken {\sl by mass terms},
the quadratic divergence does not reappear!
All we need in order to see this is dimensional analysis.
Suppose we take a supersymmetric theory and change the scalar mass 
(squared)~\footnote{For scalar fields, the physical parameter is the mass-squared. This is what
appears in the Lagrangian.}
\beq
{m_0}_{scalar}^2 = {m_0}_{fermion}^2 + \mtilde^2
\eeq
where $\mtilde^2$ is some constant.
Will there be a quadratic divergence in the scalar mass?
\beq
\delta m_{scalar}^2 = \# \Lambda^2 + \# {m_0}_{scalar}^2 
\log\frac{{m_0}_{scalar}^2}{\Lambda^2} ~~~~??
\eeq
No.
For $\mtilde^2=0$,  supersymmetry is restored, and therefore there shouldn't be a
quadratic divergence.
So the $\Lambda^2$ term (which is the quadratic divergence) must be proportional to $\mtilde^2$.
But there's nothing we can write in perturbation theory
that would have the correct dimension.

We conclude that, if supersymmetry is broken by
\beq
{m_0}_{scalar}^2 \neq {m_0}_{fermion}^2 
\eeq
{\bf the scalar mass-squared has only log divergences}.
In other words,
the supersymmetry breaking (given by the fact that the scalar mass is different from 
the fermion mass) does not spoil the cancellation
of the quadratic divergence.
This type of breaking is called {\bf soft}-supersymmetry breaking.
(This is what we have in the Minimal Supersymmetric Standard Model (MSSM).)

%

Parenthetically, we note that one can also have
{\bf hard} supersymmetry breaking.  
Take a supersymmetric theory, and change some dimension-{\sl less} number,
eg, the coupling of the boson compared to the coupling of  the fermion.
This will
reintroduce the quadratic  divergences.

We derived all these results based on dimensional analysis.
Now let's see them concretely.
To get something non-trivial we must add interactions.
Let's go back to our simple theory,
\beqa
{\cal L}&=& 
\partial^\mu\phi_+^* \,\partial_\mu\phi_+ 
+ \psi_+^\dagger i\bar\sigma^\mu \partial_\mu \psi_+\\
&+&\partial^\mu\phi_-^* \,\partial_\mu\phi_- +
\psi_-^\dagger i\bar\sigma^\mu \partial_\mu \psi_-\\
&-&m^2 \left\vert \phi_- \right\vert^2 
-m^2 \left\vert \phi_+ \right\vert^2 
-m ( \psi_+^T\varep \psi_- + {\rm hc})
\eeqa  
Our two fermions look like the two pieces of an electron or a quark.
For example, you can think of
$\psi_-$ as the SM SU(2)-doublet quark,
and of the
$\psi_+$ as the SM SU(2)-singlet quark.
To get interactions, 
let's add a complex scalar  $h$,
with the ``Yukawa'' interaction:
\beq
\delta {\cal L}= - y\, h\, \psi_+^T \varep \psi_- +{\rm hc}
\eeq
here $y$ is a coupling.
To make a supersymmetric theory we also need 
a  (L) fermion $\tilde h$
(their supersymmetry transformations are just like $\phi_+$ and $\psi_+$)
\footnote{
If $h$ and $\tilde h$ remind you of the Higgs and Higgsino that's great, but
here they have nothing to do with generating mass, we are just interested 
in the interactions.}.
Finally, just for simplicity, let's set $m=0$.

It's easy to see that if we just add this Yukawa interaction, the Lagrangian
is not invariant under supersymmetry.
So we must add more interactions,
\beqa
{\cal L}&=& 
\partial^\mu\phi_+^* \,\partial_\mu\phi_+ 
+\partial^\mu\phi_-^* \,\partial_\mu\phi_- +
\partial^\mu h^* \,\partial_\mu h \nonumber\\
&+& \psi_+^\dagger i\bar\sigma^\mu \partial_\mu \psi_+
+ \psi_-^\dagger i\bar\sigma^\mu \partial_\mu \psi_-
+ \tilde h^\dagger i\bar\sigma^\mu \partial_\mu \tilde h
\nonumber\\
&+& {\cal L}_{int}\,,
\eeqa  
with
\beqa
{\cal L}_{int} = &-& y\, \large(h \psi_+^T \varep \psi_- 
+\phi_+\tilde h^T \varep \psi_- + \phi_-\tilde h^T \varep \psi_+ 
+{\rm hc}\large)\nonumber\\
&-&
\vert y\vert^2 \large[
\vert \phi_+\vert^2 \, \vert \phi_-\vert^2
+\vert h\vert^2 \, \vert \phi_-\vert^2
+\vert h\vert^2 \, \vert \phi_+\vert^2
\large]\,.
\eeqa

%


Now that we have an interacting supersymmetric theory,
we are ready to consider the UV divergence in the scalar mass-squared.
Consider $\delta m_h^2$. It gets contributions from 
a $\phi_+$ loop, a $\phi_-$ loop and a fermion loop.



%

To calculate the fermion loop, let's convert to Dirac fermion language,
\beq
y\, h \psi_+^T \varep \psi_- +{\rm hc} = y h \bar\psi P_L \psi + {\rm hc}\,.
\eeq
So the fermion loop is 
 \beq
-\vert y\vert^2 \int \frac{d^4p}{(2\pi)^4} \tr P_L \frac{i}{\slam p} P_R
\frac{i}{\slam p} = 2 \vert y\vert^2 \int \frac{d^4p}{(2\pi)^4}\frac{1}{p^2}\,.
\eeq
(In the MSSM, the analog of this is the top contribution to the Higgs mass).

The boson loop is,
\beq
2\times i\vert y\vert^2 \int \frac{d^4p}{(2\pi)^4} \frac{i}{p^2}
 = -2 \vert y\vert^2 \int \frac{d^4p}{(2\pi)^4}\frac{1}{p^2}\,,
\eeq
(in the MSSM, the analog of this is the stop contribution to the Higgs mass).
%


Before we argued that the cancellation is not spoiled by soft supersymmetry
breaking.
Let's see this in this example.
Suppose we change the $\phi_\pm$ masses-squared to
$\mtilde^2_\pm$.
Indeed there is no quadratic divergence,
\beqa
\delta m^2_h &\propto& \vert y\vert^2 \int \frac{d^4p}{(2\pi)^4} 
\left[\frac{2}{p^2} - \frac{1}{p^2-\mtilde^2_+}
- \frac{1}{p^2-\mtilde^2_-}
 \right]\\ 
&=&
\vert y\vert^2 \, \mtilde^2_1\,
\int \frac{d^4p}{(2\pi)^4} 
 \,\frac1{p^2(p^2-\mtilde^2_+)}
+ (\mtilde^2_+ \to  \mtilde^2_-)\,.
\nonumber
\eeqa
We see that
when supersymmetry is softly broken,
the scalar mass squared is log divergent, and
the divergence is proportional to the supersymmetry breaking $\mtilde^2$.
In contrast to ``hard'' supersymmetry  breaking: if we change one of the 4-scalar couplings
from $\vert y\vert^2$, the  quadratic divergence is not cancelled.

%


We now know a lot of supersymmetry basics.
Let's recap and add some language:
\begin{itemize}
\item 
Supersymmetry is an extension of the Poincare symmetry: it's a spacetime symmetry.
\item
The basic supersymmetry ``module'' we know is
a complex scalar $+$ a 2-component spinor {\sl of the same mass}.
eg,
\beq
(\phi_+, \psi_+)
\eeq
These transform into each other under supersymmetry.
Together they form a representation, or a multiplet of supersymmetry.
For obvious reasons, we call this the ``chiral supermultiplet''.
\item
The number of fermionic dof's equals the number of bosonic dof's.
(This is true generally.)
\item
Supersymmetry dictates not just the field content but also the interactions.
The couplings of fermions, bosons of the same supermultiplets are related.
(Again, this is true generally.)
Starting from a scalar1--fermion2--fermion3 vertex,
supersymmetry requires also a
fermion1--scalar2--fermion3 vertex and a fermion1--fermion2--scalar3 vertex
all with same coupling, as well as
4-scalar vertices (with the {\bf same} coupling squared).
\end{itemize}
You see that the structure of supersymmetric theories is very constrained,
and that as a result it's less divergent.
(This is the {\sl real} reason theorists like supersymmetry,
it's easier.. In fact, the more supersymmetry, the easier it gets. 
There are less divergences, more constraints, one can calculate many things,
even at strong coupling.
By the time you get to maximal supersymmetry in 4d you have 
a finite, scale invariant theory.)

We also know a great deal about supersymmetry breaking, so let's summarize that too.
\begin{itemize}
\item
With unbroken supersymmetry the  vacuum energy is zero.
Thus the vacuum energy is an order parameter for supersymmetry breaking,
and supersymmetry breaking always involves a scale $E_{\rm vac}$.
\item Supersymmetry (breaking) and UV divergences:
With  unbroken supersymmetry we have only log divergences.
Even in the presence of
soft supersymmetry breaking (ie, supersymmetry is broken by dimensionful quantities only),
there are  only log divergences.
In contrast,
hard supersymmetry breaking (ie when pure numbers, 
such as couplings, break supersymmetry) reintroduces
quadratic divergences, so it's not that interesting from the point of view of
the fine-tuning problem.
\end{itemize}

Let's pause and talk about language.
This will be useful when we supersymmetrize the SM.
Our simple example of eqn~(\ref{toy}) has two chiral supermultiplets, 
each contains one complex scalar and one  L-handed fermion,
\beq
(\phi_+ ~~~ \psi_+)\,, ~~~~ (\phi_-~~~~~\psi_-)
\eeq
In the SM each fermion, eg the top quark, comes from a fusion
of 2 Weyl fermions:
one originating from an SU(2) doublet and the other from an SU(2) singlet.
These are the analogs of $\psi_+$, $\psi_-$.
When we supersymmetrize the SM we must add two scalars (the stops, or top squarks)
 these are the analogs of $\phi_+$, $\phi_-$.
One often refers to the doublet and singlet fermions as
``L-handed'' and ``R-handed''. 
This is bad language (remember we can always write a left handed spinor using 
a right-handed spinor).
If we used this bad language anyway, we could call our fermions
$\psi_L$, $\psi_R$,
and the accompanying scalars:
$\phi_L$, $\phi_R$.
This is why you hear people talk about 
the stop-left and stop-right, or left squarks and right squarks.
Of course the stops are scalars, and have no chirality, but
the names just refer to their fermionic partners.

\subsection{Spontaneous supersymmetry breaking: the vacuum energy, UV divergences}
If supersymmetry is realized in Nature it's realized as a broken symmetry.
We already saw that even explicit (soft) supersymmetry breaking can be
powerful. 
But the picture we had is not very satisfying: we don't want to put in the parameter
$\mtilde^2$ by hand.
We want it to be generated by the theory itself:
we want the theory to break supersymmetry {\sl spontaneously}.

We also saw that with unbroken supersymmetry the vacuum energy vanishes, and
the potential $V\geq0$.
Thus,
supersymmetry is unbroken if there are  solution(s) 
of the EOMs with $V=0$.
Recall that this followed from
\beq
\langle0\vert\left\{
{\mathrm{SUSY, SUSY}}
\right\}
 \vert0\rangle\propto \langle0\vert H\vert0\rangle
\eeq
and with unbroken SUSY,
\beq
{\mathrm{SUSY}} \vert0\rangle=0
\eeq

However, if supersymmetry is {\sl spontaneously} broken:
\beq
{\mathrm{SUSY}} \vert0\rangle\neq0
\eeq
and
the ground state has nonzero (positive) energy!

In the SM, the only scalar is the Higgs, so the only potential is
the Higgs potential, and we're not that used to thinking about 
scalar potentials.
But in supersymmetric theories, fermions are always accompanied by scalars and
any fermion interaction results in a scalar potential,
eg
\beq
V(\phi_+, \phi_-,h)=\vert y\vert^2 \left[
\vert \phi_+\vert^2 \, \vert \phi_-\vert^2
+\vert h\vert^2 \, \vert \phi_-\vert^2
+\vert h\vert^2 \, \vert \phi_+\vert^2\right]
\eeq
Note that indeed: $V\geq0$---unbroken supersymmetry.

To break supersymmetry spontaneously all we need is to find a supersymmetric 
theory with a potential which is always {\sl above} zero.
So we need a scale.
Classically, we can just put in scale by hand.
This brings us to the simplest example of spontaneous supersymmetry breaking.

\subsubsection{The O'Raifeartaigh model}
The simplest supersymmetric theory with chiral supermultiplets
that breaks supersymmetry spontaneously 
has three chiral supermultiplets,
\beq
(\phi, \psi)\,,~~~~(\phi_1, \psi_1)\,,~~~~~~(\phi_2, \psi_2)
\eeq
and two mass parameters.
We will only write the scalar potential\footnote{The kinetic terms and 
fermion-fermion-scalar interactions are there too, but there's nothing
instructive in them at this point.},
\beq
V= \vert  y\phi_1^2-f\vert^2 + m^2 \vert \phi_1\vert^2 +
\vert 2\phi_1\phi +m\phi_2\vert^2
\eeq
Here $m$ is a mass, $f$ has dimension mass$^2$, 
and $y$ is a dimensionless coupling.
It is  easy to see that there is no supersymmetric minimum.
The first two terms cannot vanish simultaneously.
Supersymmetry is broken! 
Note that we need $f\neq0$ for that (we must push some field away from the origin)
as well as $m\neq0$.

Finding the ground state requires more effort.
Let's assume $f < m^2/(2y)$. 
The ground state is at $\phi_1=\phi_2=0$ with $\phi$ arbitrary
($\phi$ is a flat direction of the potential),
\beq
V_0= \vert f\vert^2
\eeq
Expanding around the VEVs, one finds the following spectrum:
One massless Weyl fermion, one Dirac fermion of mass $m$,
and several real bosons of which two are massless,
two have  mass $m$, one has mass $\sqrt{m^2 +2yf}$,
and  one $\sqrt{m^2 -2yf}$. 
Indeed, for $f=0$ supersymmetry is restored, and the fermions and bosons become degenerate.

Why are there massless bosons in the spectrum?
Recall that $\phi$ is arbitrary, it's a flat direction
(2 real dof's).
Why is there a massless Weyl fermion?
Normally a spontaneously broken global symmetry implies the existence of a massless 
Goldstone boson (or pion).
Here we have  spontaneously broken {\sl super}symmetry, which is a ``fermionic'' symmetry,
so we have a massless Goldstone {\sl fermion}.
Since supersymmetry is broken spontaneously, the supersymmetry generator does 
not annihilate the vacuum,
\beq
{\mathrm{SUSY}} \vert0\rangle\neq0
\eeq
where SUSY stands  for the supersymmetry generator.
Since this generator carries a spinor index, 
this state is a fermion state, which is precisely the Golsdtone fermion 
(sometimes called a Goldstino).

Recall that we needed a scale, or a dimensionful parameter in order to break supersymmetry. 
Above we simply put it in by hand.
But suppose we started with no scale in the Lagrangian. Then classically, supersymmetry
would remain unbroken. 
This suggest that, since no scale can be generated perturbatively, if supersymmetry 
is unbroken at the tree-level, it remains unbroken to all orders in perturbation theory.
This is actually true, and it is a very powerful result.
It's a consequence of the constrained structure of supersymmetry.
So if supersymmetry is unbroken at tree level, it can only be broken by non-perturbative
effects, with a scale that's generated dynamically, just like the QCD scale,
\beq
\Lambda = M_{UV} \, \exp\left(\frac{-8\pi^2}{b g^2}\right)
\eeq
which is exponentially suppressed compared to the cutoff scale.

This type of supersymmetry breaking is called, for obvious reasons,
{\sl dynamical} supersymmetry breaking.
We will come back to this when we discuss the standard model.
It leads to a beautiful scenario: 
The supersymmetry breaking scale can naturally be 16 or so orders
of magnitude below the Planck scale.

\subsection{The chiral and vector multiplets}
We have the chiral supermultiplet:
\beq
(\phi, \psi)
\eeq
with $\phi$ a complex scalar, $\psi$ a 2-component fermion.
In the real world we also have spin-1 gauge bosons, $A_\mu^a$, where $a$ denotes the 
gauge group index.
So in order to supersymmetrize the SM we also need
vector supermultiplets,
\beq
(A_\mu^a, \lambda^a)
\eeq
namely, a gauge field + a ``gaugino''.
On-shell $A_\mu^a$ has 2 dof's (2 physical transverse polarizations), so $\lambda^a$ is 
a 2-component spinor.
$A_\mu^a$ is real, so if we want to write $\lambda^a$ as a 4-component spinor
it must be a Majorana 
spinor\footnote{
As opposed to the Dirac fermion which consists of two distinct 2-component fermions,
\beq
 \psi=
\begin{pmatrix}
\psi_-\\
-\varep \psi_+^* 
\end{pmatrix}
\eeq.},
\beqa
\begin{pmatrix}
\lambda\\
-\varep\lambda^*
\end{pmatrix}
\eeqa
Under a supersymmetry transformation,
 $A_\mu^a$ ~$\lambda^a$ transform into each other,
and we can construct supersymmetric Lagrangians for them
as we did for the chiral supermultiplet.

\subsection{Supersymmetric Lagrangians}
We now have the gauge module (gauge field + gaugino)
and the chiral module (scalar + fermion).
What are the Lagrangians we can write down?
With a theory of such a constrained structure, you expect to have many 
limitations.
Indeed, all the theories we can write down are encoded by two functions,
the K\"ahler potential (K), which gives the kinetic and gauge 
interactions\footnote{
As we will see, there is no freedom there at the level of 4d terms
so we won't even write it down.},
and
the superpotential (W), which gives the non-gauge (Yukawa like) interactions
of chiral fields.

Let's start with the gauge part: after all, gauge interactions are
 almost all we measure.

\subsection{A pure supersymmetric gauge theory}
We want a gauge-invariant supersymmetric Lagrangian for
\beq
(A_\mu^a, \lambda^a)
\eeq
Gauge symmetry (and supersymmetry) determine it completely up to higher-dimension terms.
It is
\beq
{\cal L}_{gauge} = -\frac14 F^{a\mu\nu}F^a_{\mu\nu} 
+ \lambda^{a\dagger} i\bar\sigma\cdot {\cal D} \lambda^a
\eeq

{\bf Exercise:}
Check that this Lagrangian is supersymmetric.

\subsection{A supersymmetric theory with matter fields and only gauge interactions}
We also want to couple ``matter fields'' to the gauge field.
So we add our chiral modules $(\phi_i,\psi_i)$.
Here too, there is no freedom because of gauge symmetry plus supersymmetry,
\beqa
{\cal L}&=& {\cal L}_{gauge} +
{\cal D}^\mu\phi_i^* \,{\cal D}_\mu\phi_i 
+ \psi_i^\dagger i\bar\sigma^\mu {\cal D}_\mu \psi_i 
\no\\
&-&\sqrt2 g\, ( \phi_i^* \lambda^{a T} T^a \varep \psi_i 
- \psi_i^\dagger\varep \lambda^{a*} T^a \phi_i) 
- \frac12\, D^a D^a
\eeqa
where
\beq
D^a = -g  \phi_i^\dagger T^a \phi_i 
\eeq
As in the chiral theory, supersymmetry dictates a ``new'' coupling.
In non-supersymmetric theories we have a coupling\\
\centerline{gauge field---fermion---fermion}
\noindent
Now we also have,\\
\centerline{gaugino---fermion---scalar,}
and of course there's also \\
\centerline{gauge field---scalar---scalar.}

In addition, there is a scalar potential with a 4-scalar interaction,
\beq
V= \frac12 D^a D^a
\eeq
with
\beq
D^a = -g  \phi_i^\dagger T^a \phi_i 
\eeq
This is a quartic scalar potential,
but the quartic coupling isn't arbitrary,  it's the gauge coupling.
This will be very important when we discuss the Higgs!

Note what happened here. 
Starting from a non-supersymmetric gauge theory with a gauge field $A_\mu^a$
and a fermion $\psi$, and an 
interaction $g\, A_\mu-\psi-\psi$,
when we supersymmetrize the theory, the 
field content is gauge field + gaugino, fermion + scalar.
The interactions are $A_\mu-\phi-\phi$ (nothing new, $\phi$ is charged),
but also, $\lambda-\phi-\psi$  (gaugino-scalar-fermion)
all with same coupling $g$.
In addition there is a 4-scalar interaction, with coupling $g^2$.
We had no freedom in the process. The field content and couplings of the supersymmetric theory
were dictated by (i) the original non-supersymmetric theory we started from 
(ii) the gauge symmetry (iii) supersymmetry.

\subsection{Yukawa like interactions}
We also want Yukawa-like interactions of just the chiral scalars and fermions
$(\phi_i,\psi_i)$.
There is a simple recipe for writing down the most general supersymmetric 
interaction Lagrangian.
Choose an analytic function $W(\phi_1,\ldots,\phi_n)$---the ``superpotential''.
Analytic means that $W$ is not a function of the conjugate fields
(no daggers!).
All the allowed interactions are given by
\beq
{\cal L_{\rm int}}= 
-\frac12\,\frac {\partial^2 W}{\partial \phi_i\partial\phi_j}\, 
\psi_i^T \varep \psi_j +{\rm hc}
-\sum_i \left\vert
F_i
\right\vert^2
\eeq
where
\beq
F_i^* =-\frac{\partial W}{\partial\phi_i}
\eeq
This Lagrangian is guaranteed to be supersymmetric! (There is an elegant way to see this.)

Let's write our previous examples in this language.
Start with the theory containing $h$, $\phi_+$, $\phi_-$:
Take
\beq
W= y\, h\, \phi_+\, \phi_-
\eeq
so 
\beqa
F_h^*&=&-\frac{\partial W}{\partial h} = y \phi_+ \phi_-\no\\
\frac{\partial^2 W}{\partial h\,\partial\phi_+} &=& y \phi_-
\eeqa
and similarly for the remaining fields.
We indeed recover
\beqa
{\cal L}_{int} = &-& y\, \large(h\, \psi_+^T \varep \psi_- 
+\phi_+\,\tilde h^T \varep \psi_- + \phi_-\,\tilde h^T \varep \psi_+ 
+{\rm hc}\large) \no\\
&-&
\vert y\vert^2 \large(
\vert \phi_+\vert^2 \, \vert \phi_-\vert^2
+\vert h\vert^2 \, \vert \phi_-\vert^2
+\vert h\vert^2 \, \vert \phi_+\vert^2
\large)
\eeqa

{\bf Exercise:}
Check that the massive theory with $\phi_\pm$
is obtained from 
\beq
W= m\phi_+ \phi_-
\eeq

{\bf Exercise:}
Check that the O`Raifeartaigh model
is obtained from 
\beq
W= \phi\,(y\phi_1^2-f) +m \phi_1\phi_2
\eeq

\subsection{R-symmetry}
We now know how to write the most general supersymmetric theory
in 4d (with minimal supersymmetry).
Note that the theory has a global U(1) symmetry.
Under this U(1), the gauge boson has charge 0, the
gaugino has charge +1, the chiral fermion has charge 0 and the scalar $-1$.
Alternatively we could take the fermion to have
charge $-1$  and the scalar to be 0.
This is called a U(1)$_R$ symmetry. It does not commute with supersymmetry:
members of the same supermultiplet have different charges.
  
This symmetry (or its remanent) is crucial in LHC supersymmetry searches!

\subsection{$F$-terms and $D$-terms}
In writing the theory,
we defined $F$-terms and $D$-terms.
For each vector multiplet $(\lambda^a, A_\mu^a)$ we have  
\beq
D^a= -g  \phi_i^\dagger T^a \phi_i ~~~~~~~({\rm dim-2})
\eeq
(to which all the charged scalars contribute).

For each chiral multiplet $(\phi_i, \psi_i)$
we have a
\beq
F_i^* =-\frac{\partial W}{\partial\phi_i}~~~~~({\rm dim-2})
\eeq
and the scalar potential is
\beq
V = F_i^* F_i +\frac12\, D^a D^a
\eeq
With this language, we can revisit supersymmetry breaking.
First we immediately see that indeed $V\geq0$.
Second, 
supersymmetry is broken if some $F_i\neq$ or $D^a\neq0$.
So the $F_i$'s and $D^a$'s are order parameters for supersymmetry breaking.

\subsection{Local supersymmetry: the gravitino mass and couplings}
So far we thought about supersymmetry as a global symmetry.
Translations and Lorentz transformations are however {\sl local} symmetries. 
The ``gauge theory'' of local spacetime symmetry is gravity.
We therefore have no choice: supersymmetry is a local symmetry too.
The theory of local (spacetime and) supersymmetry is called supergravity. 
The spin-2 graviton must have a supersymmetric partner,  the gravitino, 
which has spin-3/2. 
Since supersymmetry is broken the gravitino should get mass.

If you're only interested in collider experiments, should you care about this?
Normally, the effects of gravity are suppressed by the Planck scale
and we can forget about them when discussing HEP experiments.
However, the gravitino mass is related to a broken local symmetry 
(supersymmetry), so just as in the usual Higgs mechanism of electroweak symmetry
breaking, it gets mass by ``eating'' the Goldstone fermion.
Thus, a piece of the gravitino (the longitudinal piece), is some 
``ordinary'' field (which participates in supersymmetry breaking),
and the gravitino couplings to matter are not entirely negligible.
Furthermore, they are dictated by the supersymmetry breaking.
If supersymmetry is broken by some non-zero $F$ term,
the gravitino mass is
\beq
m_{3/2}= \# \frac{F}{M_P}
\eeq

\section{The Supersymmetrized Standard Model}\label{ssm}
\subsection{The Supersymmetrized SM: motivation and structure}
Now that we understand what supersymmetry is, we can supersymmetrize the SM.
Let's review first the motivations for doing that.
Before 2012, all fundamental particles we knew had spin-1 or spin-$1/2$.
We now have the Higgs, which is spin 0.
This is the source of the fine-tuning problem, or the naturalness problem.
Since the Higgs is spin-0, its
mass is quadratically divergent
\beq
\delta m^2 \propto \Lambda_{UV}^2\,,
\eeq
unlike fermions, whose masses are protected by the chiral symmetry as we saw,
or gauge bosons, whose masses are protected by gauge symmetry.

In the case of the Higgs mass, there are one loop corrections that are quadratically
divergent. 
The dominant one is from the top quark.
This is not a practical problem. We can calculate any physical observable by including
a counter term that cancels this divergent contribution.
Rather, the problem is of a theoretical nature.
We believe that $\Lambda_{UV}$ is a concrete physical scale,
such as the  mass scale of new fields, or the scale of new strong interactions. 
Then at the low-scale $\mu$
\beq\label{fine}
m^2(\mu)= m^2(\Lambda_{UV}) +  \#\,\Lambda_{UV}^2
\eeq
$m^2(\Lambda_{UV})$ determined by the full UV theory, and the number
is $\#$ determined by the SM.
We know the LHS of eqn.~(\ref{fine}): $m^2\sim 100$~GeV$^2$.
So if $\Lambda_{UV}$ is the Planck scale $\sim10^{18}$~GeV 
we need
$m^2(\Lambda_{UV})\sim 10^{36}$~GeV$^2$
and the two terms on the RHS must be tuned to 32 orders of magnitude.. 
Such dramatic tunings do not seem natural.
In general, for a cutoff scale $\Lambda_{UV}$,
the parameters of the two theories must be  tuned to
${{\rm TeV}^2}/{\Lambda_{UV}^2}$.


As we saw above,
with supersymmetry (even softly broken),
scalar masses-squared have 
only log divergences:
\beq
m^2(\mu)= m^2(\Lambda_{UV}) \left[  1+\#\,
\log\left(\frac{m^2(\Lambda_{UV})}{\Lambda_{UV}^2}\right)
\right]
\eeq
 just as for fermions! 
The reason is that supersymmetry ties the scalar mass to the fermion mass.

The way this happens in practice is that
the quadratic divergence
from fermion loops is cancelled by the quadratic divergence from
 scalar loops.
The cutoff scale then only enters in the log, and 
$m^2(\Lambda_{UV})$ can be order (100~GeV)$^2$.
This is the main motivation for supersymmetric extensions of the SM.
There are further motivations too.
Supersymmetric extensions of the SM  often supply 
dark matter candidates, 
new sources of CP violation etc.
Finally, extending space time symmetry is theoretically appealing.

So let's supersymmetrize the SM.
Each gauge field is now part of a vector supermultiplet:
for the gluon we have,
\beq
G_\mu^a \rightarrow (\tilde g^a,G_\mu^a) +D^a\,,
\eeq
where the physical fields are the gluon and the spin-1/2 gluino.
Similarly for the  $W$, 
\beq
W_\mu^I \rightarrow (\tilde w^I,W_\mu^I)+ D^I
\eeq
where the physical fields are the $W$ and the wino,
and for $B$
\beq
B_\mu \rightarrow (\tilde b, B_\mu) +D_Y\,,
\eeq
where the physical fields are the $B$ and the bino.

Each fermion is now part of a chiral supermultiplet of the form
\beq
(\phi,\psi) + F\,.
\eeq
Taking all the SM fermions 
$q,u^c,d^c,l,e^c$ 
to be L-fermions, we have
\beq
q \rightarrow (\tilde q, q)+F_q ~~~{\rm 
all~transforming~as}~~ (3,2)_{1/6}
\eeq
with the
physical fields being the (doublet) quark $q$ and a spin-0 squark $\tilde q$.
Similarly,
\beq
u^c \rightarrow (\tilde u^c, u^c)+F_u~~~{\rm 
all~transforming~as}~~ (\bar3,1)_{-2/3}
\eeq
with the 
physical fields being the (singlet) up-quark $u^c$ + up squark $\tilde u^c$,
\beq
d^c \rightarrow (\tilde d^c, d^c)+ F_d~~~{\rm 
all~transforming~as}~~ (\bar3,1)_{1/3}
\eeq
with the
physical fields being the (singlet) down-quark $d^c$ + down squark $\tilde d^c$,
\beq
l \rightarrow (\tilde l, l)+ F_l~~~{\rm 
all~transforming~as}~~ (1,2)_{-1/2}
\eeq
with the 
physical fields being the (doublet) lepton $l$ + a slepton $\tilde l$,
and finally
\beq
e^c \rightarrow (\tilde e^c, e^c)+F_e~~~{\rm 
all~transforming~as}~~ (1,1)_1
\eeq
with the
physical fields being the (singlet) lepton $e^c$ + a slepton $\tilde e^c$.

Once electroweak symmetry is broken the doublets split:
\beqa
q=
\begin{pmatrix}
u\\d
\end{pmatrix}
~~~~~
\tilde q=
\begin{pmatrix}
\tilde u\\
\tilde d
\end{pmatrix}
\eeqa
and
\beqa
l=
\begin{pmatrix}
\nu\\
l
\end{pmatrix}
~~~~~
\tilde l=
\begin{pmatrix}
\tilde \nu\\
\tilde l
\end{pmatrix}
\eeqa

Now let's move on to the interactions, staring with the 
gauge interactions.
There is nothing we have to do here. As we saw above, these interactions are 
completely dictated by supersymmetry and the gauge symmetry.
We wrote the Lagrangian for a general gauge theory in the previous lecture:
\beqa
{\cal L}&=& {\cal L}_{gauge} +
{\cal D}^\mu\phi_i^* \,{\cal D}_\mu\phi_i 
+ \psi_i^\dagger i\bar\sigma^\mu {\cal D}_\mu \psi_i 
\no\\
&-&\sqrt2 g\, ( \phi_i^* \lambda^{a T} T^a \varep \psi_i 
- \psi_i^\dagger\varep \lambda^{a*} T^a \phi_i) 
- \frac12\, D^a D^a
\eeqa
where
\beq
D^a = -g  \phi_i^\dagger T^a \phi_i 
\eeq
Applying this to the SM,
\beq
\psi_i= q_i, u^c_i, d^c_i, l_i, e^c_i ~~~~
\phi_i= \tilde q_i, \tilde u^c_i, \tilde d^c_i, \tilde l_i, \tilde e^c_i 
\eeq
The covariant derivatives now contain the SU(3), SU(2) and U(1) gauge fields,
$\lambda^a$ sums over the SU(3), SU(2), U(1) gauginos
\beq
\lambda^a \to \tilde g^a, \tilde w^I, \tilde b
\eeq
and there are $D$ terms for SU(3), SU(2), U(1)
\beq
D^a \to D^a, D^I, D_Y
\eeq
In addition there is of course the pure gauge Lagrangian that I didn't write
(we saw it in the previous lecture).


The Lagrangian above contains the scalar potential,
\beq
V= \frac12 \,D^a D^a +\frac12\, D^I D^I + \frac12 \, D_Y D_Y
\eeq
where
for SU(3): (recall $T_{\bar3} =- T_3^*$ and we will write things in terms of
the fundamental generators)
\beq
D^a = g_3\, (\tilde q^\dagger T^a \tilde q - \tilde u^{c\dagger} T^{a*} u^c 
-\tilde d^{c\dagger} T^{a*} u^c) 
\eeq
similarly for the SU(2) and
\beq
D_Y = g_Y \,\sum_i Y_i \tilde f_i^\dagger  \tilde f_i 
\eeq
We see that we get 4-scalar interactions with the quartic couplings equal to the gauge 
couplings.

Again we emphasize that there was no freedom so far, and no new parameters.
We also didn't put in the Higgs field yet, so let's do this now.
The SM Higgs is a complex scalar, so it must be part of a chiral module
\beq
H \rightarrow (H, \tilde H)+F_H~~~{\rm 
all~transforming~as}~~ (1,2)_{-1/2}
\eeq
We immediately see a problem (in fact, many problems, which are all related):
First, there is a problem with having a single 
Higgs {\sl scalar}.
We want the Higgs (and {\bf only} the Higgs) to get a VEV.
However, the Higgs is charged under SU(2), U(1), so its VEV gives rise to
nonzero $D$ terms:
\beq
V\sim D^I D^I + D_Y^2
\eeq
where 
\beq
D^I = g_2\,\langle H^\dagger\rangle T^I \langle H\rangle ~~~ 
D_Y = g_1\,\frac12\,\langle H\rangle^\dagger  \langle H\rangle
\eeq
that is, EWSB implies supersymmetry breaking! 
You might think this is good, but it's not (for many reasons).
For one,
the non-zero D-terms would generate masses for the squarks and sleptons.
Consider $D_Y$ for example:
\beq
D_Y = \frac12 v^2 + \sum_i Y_i \vert\tilde f_i\vert^2
\eeq
where $\tilde f$ sums over all squarks, sleptons and $Y_i$ is their hypercharge.
Recall the scalar potential $V\sim D^2$. Therefore
some of the squarks will get {\sl negative} masses-squared of order $v^2$.
This is a disaster: SU(3) and  EM are broken at $v$!
The solution is to add a second Higgs scalar,  {\sl with opposite charges}.
The two Higgs scalars can then get equal VEVs with all $<D>=0$.


A second problem is that $\tilde H$ is a Weyl fermion.
 If this is all there is, we will have a massless fermion 
around---the Higgsino.
 In the presence of massless fermions, gauge symmetries can become anomalous,
that is, the gauge symmetry can be broken at the loop level.
In the SM, the fermion representations and charges are such that there are no anomalies.
Before discussing the Higgs, we
only added scalars to the SM (squarks and sleptons, known collectively as sfermions).
These are harmless from the point of view of anomalies.
We also added gauginos. These are fermions, but they are adjoint fermions, which
don't generate any anomalies (essentially because the adjoint is a real representation).
In contrast,
the Higgsino $\tilde H$ is a massless fermion which is a doublet
of SU(2) and charged under U(1)$_Y$.
The simplest way to cancel the anomaly is to add a second Higgsino
in the conjugate representation.
So we must add a second Higgs field with conjugate quantum numbers.
When we consider interactions, we will see other reasons why
we must do this.

We will call the SM Higgs $H_D$ and the new Higgs $H_U$.
Thus,
\beq
H_D \rightarrow (H_D, \tilde H_D) +F_{HD}~~~{\rm 
all~transforming~as}~~ (1,2)_{-1/2}\,,
\eeq
and we also add,
\beq
H_U \rightarrow (H_U, \tilde H_U)+ F_{HU}~~~{\rm 
all~transforming~as}~~ (1,2)_{1/2}\,,
\eeq
and in the limit of unbroken supersymmetry,
\beq
\langle H_U\rangle=\langle H_D\rangle \,.
\eeq

In the SM we add a quartic potential for the Higgs field,
\beq
\lambda (H^\dagger H)^2\,.
\eeq
Here there is quartic potential built in,  coming from the $D$ terms.
This potential will not necessarily give mass to the physical Higgs.

We now turn to the Yukawa couplings.
In the SM we have Higgs-fermion-fermion Yukawa couplings.
Consider the down-quark Yukawa first
\beq
y_D H_D q^T \varep d^c ~~~({\rm Higgs~ -~ quark-~ quark})\,,
\eeq
as we saw above, with supersymmetry, this must be accompanied by
\beq
+ y_D\, (\tilde q\tilde H_D^T \varep d^c +\tilde d^c\tilde H_D^T \varep q)
~~~({\rm squark~ -~ Higgsino-~ quark})\no
\no
\eeq
all coming from the superpotential
\beq
W_D= y_D H_D q d^c\,.
\eeq

Similarly for the lepton Yukawa,
\beqa
W_l&=& y_l H_D l e^c \to\\ 
{\cal L}_l&=& 
y_l ( H_D l^T \varep e^c +
\tilde l\tilde H_D^T \varep e^c +  \tilde e^c\tilde H_D^T \varep l +{\rm hc})\\
&~& {\rm Higgs-lepton-lepton} \no\\
&+&{\rm slepton-Higgsino-lepton}
\eeqa

What about the up Yukawa?
We need, 
\beq
{\rm (Higgs)} q^T \varep u^c 
\eeq
This coupling must come from a superpotential,
\beq
{\rm (Higgs)} q  u^c
\eeq
In the SM (Higgs)$=H_D^\dagger$.
But the superpotential is {\bf holomorphic},  no daggers are allowed.

This is the 4th  reason why we needed
a second Higgs field with the conjugate charges\footnote{All these are actually different 
aspects of the same problem.},
\beqa
W_U&=& y_U H_U q u^c \to\\ 
{\cal L}_U&=& 
y_U ( H_U q^T \varep u^c +
\tilde q\tilde H_U^T \varep u^c +  \tilde u^c\tilde H_U^T \varep q) +{\rm hc}
\eeqa

You can now see what's going on. In some sense,
holomorphy makes a scalar field ``behave like a fermion''.
In a supersymmetric theory, the interactions of scalar fields are
controlled by the superpotential, which is holomorphic.
For a fermion to get mass you need an LR coupling. 
So starting from a L-fermion you need a R-fermion, 
or another L-fermion with the opposite charge(s).
For a scalar $\phi$ to get mass in a non-supersymmetric theory:
you don't need anything else (you can just use $\phi^*$ to write a charge neutral mass term).
Not so in a supersymmetry theory: 
because you cannot use $\phi^*$, you must have another scalar with the 
opposite charge(s), just as for fermions.

To summarize, 
we have 2 Higgs fields $H_U$ and $H_D$.
The SM Yukawa couplings come from the superpotential
\beq
W= y_U H_U q u^c + y_D H_D q d^c + y_l H_D l e^c\,.
\eeq
Note again that there was no freedom here, and no new parameter.


 \subsection{R-symmetry}
Our supersymmetric Lagrangian also has a U(1)$_R$ symmetry.
Here is one possible choice of charges: 
gaugino ($-1$), 
sfermions (1),
Higgsinos (1), with all other fields, namely the SM fields, neutral.
You can easily check that the Lagrangian is invariant.



To recap, 
we wrote down the Supersymmetric Standard Model. It contains
\begin{itemize}
\item gauge bosons + (spin 1/2) gauginos,
\item fermions + (spin 0) sfermions, 
\item 2 Higgses + 2 (spin 1/2) Higgsinos
 \end{itemize}
The interactions are all dictated by the SM interactions + supersymmetry:
The new interactions are
\begin{itemize}
\item gauge-boson---scalar---scalar
\item
gauge-boson---gauge-boson---scalar---scalar
\item
gaugino-sfermion-fermion
\item gauge-boson---Higgsino---Higgsino
\item 4-scalar (all gauge invariant contributions)
 \end{itemize}
All the couplings are determined by the SM gauge couplings.
In particular, there is a quartic Higgs coupling which is proportional
to the gauge-coupling squared.

 
Furthermore, there is the Yukawa part,
which now contains\\
Higgsino---fermion---sfermion\\
with a 
coupling equal to the SM Yukawa coupling.

The Lagrangian is invariant under a 
U(1)$_R$ symmetry:
in each of the interactions, the new superpartners appear in pairs!
This is important both for  LHC production
and for DM.

 There is now
no quadratic divergence in the Higgs mass.
Each quark contribution is canceled by the corresponding 
squark contribution.
In particular 
the top loop is canceled by the L, R stops.
Similarly, the contribution from the  Higgs self coupling 
(from the $D$ term) is canceled by the Higgsinos,
and
each gauge boson contribution is canceled by the gaugino contribution.

But we now have,  
a wino degenerate with the $W$,
a selectron degenerate with the electron, etc.
Supersymmetry must be broken. 
Somehow the wino, selectron, and all the new particles should get mass.
It would be nice if the supersymmetrized SM broke supersymmetry 
spontaneously (after all we have lots of scalars with a complicated potential). But it does not, and
so we must add more fields and interactions that break supersymmetry.
These new fields must couple to the SM fields in order to generate masses
for the superpartners.


\subsection{The supersymmetrized standard model with supersymmetry-breaking
superpartner masses}


 \subsubsection{General structure}

The general structure is then

\centerline{SB {\bf{\Large-----}} SSM}

\noindent
Here SSM is the Supersymmetrized SM.                             
SB is a set of new fields and interactions such that supersymmetry is 
spontaneously broken. As a result there are
mass splittings between the bosons and fermions of the same SB multiplet.

Finally, {\bf{\Large-----}} stands for
 some coupling(s) between the SSM fields and the SB fields.
Since there are supersymmetry-breaking
mass-splittings among the SB fields, this coupling will 
generate
mass splitting between the SM fields and their superpartners,
mediating the supersymmetry breaking to the MSSM.
The mediation mechanism determines the supersymmetry-breaking terms 
in the MSSM, which in turn determine the experimental signatures
of supersymmetry.

\subsubsection{The supersymmetry-breaking terms:
what do we expect?}
Any term is allowed in the Lagrangian unless a symmetry prevents it.
Now that we broke supersymmetry,  supersymmetry breaking terms 
are allowed. 
In the matter sector,  sfermions get mass. However,
the fermions don't: they are protected by chiral symmetry.
In the gauge sector, gauginos get mass.
However, gauge bosons don't: they are protected by gauge symmetry.
In the Higgs sector, the Higgses get mass.
Higgsinos don't, they are protected by chiral symmetry.
This is a problem. We would like the gauginos to get mass,
so we will have to solve this problem. 

In addition, there are trilinear scalar terms that can appear,
such as a Higgs---squark---squark coupling, or a
Higgs---slepton---slepton coupling.
These are allowed 
by gauge symmetry, and supersymmetry is no longer there
to forbid them. These terms are called $A$-terms.

Thus the supersymmetry-breaking part of the SSM Lagrangian is:
\beqa
{\cal L}_{soft}&=& 
-\frac12\large[\tilde m_3 \tilde g^{T}\varep \tilde g +
 \tilde m_2 \tilde w^{ T}\varep \tilde w +
\tilde m_1 \tilde b^T\varep \tilde b\large ]\no\\
&-&\tilde q^* \tilde m_q^2 \tilde q  -
\tilde u^{c*} \tilde m_{uR}^2 \tilde u^c  -
\tilde d^{c*} \tilde m_{dR}^2 \tilde d^c  \no\\
&-&\tilde l^* \tilde m_l^2 \tilde l  -
\tilde e^{c*} \tilde m_{eR}^2 \tilde e^c  \\
&-& H_U^* m_{H_U}^2 H_U - H_D^* m_{H_D}^2 H_U\no\\
&-& H_U \tilde q^* A_U \tilde u^c -  H_D \tilde q^* A_U \tilde d^c
-H_D \tilde l^* A_l \tilde e^c\no\\
&-&B\mu H_U H_D\no
\eeqa
The last line is a quadratic term for the Higgs scalars. 
The line before last is the new trilinear scalar interactions, or $A$-terms.
When the Higgses get VEVs, these too will induce sfermion mass terms
(mixing L and R scalars).
Finally, $m_q^2$ etc are $3\times3$ matrices in generation space.
So are the A-terms ($A_U$ etc).

The values of the supersymmetry-breaking parameters
are determined by the SB theory and {\sl mainly} by the mediation.
You sometimes hear people criticize supersymmetric extensions
of the SM for having a hundred or so new parameters
(the parameters of ${\cal L}_{soft}$).
These are all determined however by the SB and the mediation scheme.
Often, these involve very few new parameters (only one in anomaly mediation and two
in minimal gauge mediation). 
 
Note too that
the parameters of ${\cal L}_{soft}$ are the only freedom we have,
and where all the interesting physics lies.
These parameters determine the spectrum
of squarks, sleptons, gauginos, and therefore
the way supersymmetry manifests itself in Nature.

\subsubsection{R-parity}
The gaugino masses and $A$-terms break the U(1)$_R$ symmetry
of the SSM Lagrangian. There is a discrete symmetry left however.
This remanent symmetry is called R-parity.
Under R-parity, the gauginos, sfermions, and Higgsinos are odd,
and all SM fields are even.
Thus, when we supersymmetrize the SM without adding any new 
interactions, we have a new parity symmetry.
It follows that the the lightest superpartner (LSP) is stable!

 \subsubsection{The mu-term: a supersymmetric Higgs and Higgsino mass}
Before we go on, let's discuss one remaining problem.
We have two massless Higgsinos in the theory. As we saw above,
these do not get mass from supersymmetry breaking.
So we must also include a supersymmetric mass term for them,
\beq
W= \mu H_U H_D\,.
\eeq

 \subsection{Mediating the breaking}
What can mediate supersymmetry breaking? 
What can the coupling {\bf{\Large-----}} be?
There are many possibilities.
One is gauge interactions. This is the basis of Gauge Mediated 
Supersymmetry Breaking (GMSB).
Another is  gravity. This is the basis of 
Anomaly Mediated Supersymmetry Breaking
(AMSB). 
Planck-suppressed interactions, which are also associated with gravity, 
are at the basis of ``gravity mediated supersymmetry 
breaking''\footnote{
mSUGRA or the cMSSM are ansatze of gravity mediation
with the assumption of flavor-blind soft terms.}.
Even Yukawa-like interactions can do the job.



 \subsubsection{Gauge Mediated Supersymmetry Breaking}
Gauge interactions are the ones we know best.
Therefore gauge mediation gives  full, concrete, and often fully
calculable
supersymmetric extensions of the SM. 

We can start with a toy example to illustrate how things work.
We saw the O`Raifeartaigh model,
\beq
W= \phi\,(\phi_1^2-f) +m \phi_1\phi_2\,.
\eeq
Recall that this  model breaks supersymmetry.
The spectrum of the model contains a supermultiplet
with supersymmetry-breaking mass splittings:
a fermion of mass $m$, and
scalars of masses-squared
$m^2 +2f$,
$m^2 -2f$.

Let's complicate the model slightly, by considering five fields,
$\phi$, $\phi_{1\pm}$, $\phi_{2\pm}$, with the superpotential, 
\beq\label{orafnew}
W= \phi\,(\phi_{1+}\phi_{1-}-f) +m \phi_{1+}\phi_{2-}
+m \phi_{1-}\phi_{2+}
\eeq
now the model has a $U(1)$ symmetry, under which $\phi$ has charge zero,
and $\phi_{i\pm}$ (with $i=1,2$) has charge $\pm1$.  
It is easy to see that supersymmetry is still broken.
Again we have supermultiplets with supersymmetry breaking 
splittings between fermions and bosons.
Now let's promote  the $U(1)$ symmetry to a gauge symmetry,
and identify it with hypercharge.
Another way to think about this is the following.
Add to the SM the fields
$\phi$, $\phi_{1\pm}$, $\phi_{2\pm}$ of hypercharges 0, $\pm1$,  
respectively, 
with the superpotential~(\ref{orafnew}).
Now consider a squark.
It is charged under hypercharge, so it couples to 
these split supermultiplets.
Therefore, a squark mass is generated!



Minimal Gauge Mediation Models are the simplest models of this type.
Suppose we have a supersymmetry-breaking model with
chiral supermultiplets $\Phi_i$ and $\bar \Phi_i$, $i=1,2,3$
such that
the fermions $\psi_{\Phi_i}$ and $\psi_{\bar \Phi_i}$ combine into 
a Dirac fermion of
mass $M$, 
and
the scalars have masses-squared $M^2\pm F$
(with $F<M^2$).
Now identify $i$ as an SU(3) color index.
Thus $\Phi$ is a $3$ of SU(3), $\bar \Phi$ is a $\bar3$ of SU(3).
These fields have supersymmetry-breaking masses.
The gluino talks to the $\Phi$'s directly and therefore 
gets mass at one loop. 
The squarks talk to the gluino
and therefore get mass at two loops.
We have
a gluino mass,
\beq
m_{{\tilde g}} = \# \frac{\alpha}{4\pi}\, \frac{F}{M} +{\cal O}(F^2/M^2)
\eeq
and
a squark mass-squared at two loops:
\beq
m_{{\tilde q}}= \# \frac{\alpha^2}{(4\pi)^2}\, \frac{F^2}{M^2} +{\cal O}(F^4/M^6)
\eeq
where
the numbers are group theory factors.
We can infer this form very simply:
\begin{itemize}
\item Since the masses arise at one or two loops there is the 
appropriate loop factor.
\item The masses should vanish as $F\rightarrow0$. 
\item The masses should vanish as
$M\rightarrow\infty$.
\end{itemize}
Gauge mediation is very elegant:
\begin{itemize}
\item The soft masses are determined by the gauge couplings.
\item
The squark matrices are flavor-blind 
($\propto 1_{3\times3}$ in flavor space).
\item
The gluino masses $\sim$ squark masses.
\item
The only new parameters are $F$ and $M$, and the
overall scale  is $F/M$.
If want soft masses around the TeV, $F/M\sim 100$~TeV.
\end{itemize}
The new fields $\Phi$ are the {\sl messengers} of supersymmetry breaking.
In order to give masses to all the MSSM fields we need messenger 
fields charged
under SU(3), SU(2), U(1),
eg, $N_5$ copies of 
$(3,1)_{-1/3}+(\bar3,1)_{1/3}$
and $(1,2)_{-1/2}+(1,2)_{1/2}$
(filling up a $5+\bar5$ of SU(5)).
This adds another parameter,
namely the number of messengers,
 $N_5$.

The messenger scale $M$ mainly enters through running.
The soft masses are generated at the messenger
scale.
To calculate them at the TeV we  need to include RGE effects.

The gravitino mass in these models is 
\beq
m_{3/2}= F_{eff}/M_P
\eeq
where $F_{eff}$ is the the dominant $F$ term.
Therefore,
\beq
m_{3/2}\geq \frac{F}{M_P} \sim   100~{\rm TeV}\,\frac {M}{M_P}
\eeq
and 
for  low messenger scales, the gravitino can be very light ($\sim$eV).

 Minimal gauge mediation is just a simple example.
Gauge mediation can in principle 
have a very different structure.
The only defining feature is that the soft masses are generated by
the SM gauge interactions.
Generically then,
\begin{itemize}
\item Colored superpartners (gluinos, squarks) are heavier than
non-colored (EW gauginos, sleptons..) by a factor
\beq
\frac{\alpha_3}{\alpha_2}~~~{\rm or} ~~~\frac{\alpha_3}{\alpha_1}
\eeq
\item
In particular, gaugino masses scale as
\beq
\alpha_3:\alpha_2:\alpha_1
\eeq
and the bino is the lightest gaugino.
\item
To leading order, the $A$ terms vanish at $M$.
\item The gravitino is light.
\end{itemize}


\subsubsection{Gravity Mediation}
With gauge mediation, we had to do some real work:
add new fields, make sure they get supersymmetry-breaking masses,
couple them to the MSSM.
But supersymmetry breaking is one place where we can expect a free 
lunch.
Imagine we have, in addition to the SM, some SB fields,
eg, the O'Raifeartaigh model.
Since supersymmetry is a space-time symmetry, the SM fields should
know this automatically.
We would expect soft terms to be generated, suppressed by $M_P$.
This is known as ``gravity mediation''.
We will discuss first the purest form of gravity mediation: anomaly
mediation, 
and then what's commonly referred to as gravity mediation.


 {\bf Anomaly mediation:}
We assume that supersymmetry is broken by some fields that have
{\sl no coupling} to the SM. These fields are called the ``hidden 
sector''.
The gravitino gets mass $m_{3/2}$. 
Would the SSM ``know'' about supersymmetry breaking?
Yes: at the quantum level, it's not scale-invariant:
all the couplings (gauge, Yukawa) run---they are scale dependent.
Therefore they are sensitive to the supersymmetry-breaking 
gravitino mass, and
{\bf all} the soft terms are generated.
The gaugino masses are given by,
\beq
m_{1/2}= b\frac{\alpha}{4\pi}\, m_{3/2}
\eeq
where
$\alpha$ is the appropriate gauge coupling and
$b$ is the beta-function coefficient.
Thus
for SU(3) $b=3$, for SU(2) $b=-1$ and for U(1) $b=-33/5$.

 Sfermions get masses proportional to their anomalous dimensions:
\beq
m_0^2 \sim \frac1{16\pi^2}(y^4 - y^2 g^2 + b g^4) \, m_{3/2}^2 
\eeq
For the first and second generation sfermions, 
we can neglect the Yukawas and,
\beq
m_0^2 \sim \frac{g^4}{16\pi^2}\, b \, m_{3/2}^2 
\eeq
$A$ terms are generated too, proportional to the beta functions of the
appropriate Yukawa.

This is amazing:
These contributions to the soft terms are {\bf always there}.
All the soft terms are determined by just the SM couplings
with one new parameter, the gravitino mass.
It seems too good to be true. Indeed,
while SU(3) is asymptotically free and $b_3>0$,
SU(2), U(1) are not, $b_2, b_1<0$.
Therefore the sleptons are tachyonic.
There are various solutions to this problem,
but the gaugino masses are fairly robust,
\beq
m_{\tilde w}: m_{\tilde b}: m_{\tilde g}: m_{3/2} \sim 1: 3.3 : 10 :370 
\eeq
In this scenario,
the
wino is the LSP.
Note that
the gravitino is roughly a loop factor heavier than the SM 
superpartners.


 {\bf Gravity mediation: mediation by Planck suppressed operators:}
Let's return to our basic setup.
The SSM is the supersymmetric standard model.
SB contains new fields and interactions that break supersymmetry
(the ``hidden sector'').
Generically, we expect to have higher-dimension operators,
suppressed by $M_P$, that
couple the SB fields and the SSM fields.
Supersymmetry breaking leads to non-zero $F$ terms (or $D$ terms)
for the SB fields,
so the higher-dimension operators coupling the two sectors will 
generate supersymmetry-breaking terms in the SSM,
with sfermion mass from
\beq
\frac{\vert F\vert^2}{M_P^2} \tilde f^\dagger \tilde f
\eeq
and
gaugino masses from
\beq
\frac{\vert F\vert}{M_P} \lambda^T\varep\lambda
\eeq
You can think of these as mediated by tree-level exchange of Planck-scale
fields.


Unlike in the previous two schemes, here we don't know the 
order-one coefficients.
Consider for example the doublet-squarks.
Their mass terms are,
\beq
c_{ij}\, \frac{\vert F\vert^2}{M_P^2} \tilde q_i^\dagger \tilde q_j
\eeq
where $c_{ij}$ are order-one coefficients.
Thus,
\beq
(m_{\tilde q}^2)_{ij} = c_{ij}\, m_0^2
~~~{\rm where}~~~
m_0\equiv \frac{\vert F\vert}{M_P} 
\eeq
In ``minimal sugra'', or the cMSSM one {\bf assumes}
\beq
c_{ij}=\delta_{ij}\,.
\eeq
It is not easy to justify this: the Yukawas are presumably generated
at this high scale, so there are flavor-dependent couplings 
in the theory.

Including the 
running to low scales,
\beq
\frac{d}{dt} m_{1/2}\propto \frac{\alpha}{4\pi}\, m_{1/2}
\eeq
we find that
starting from a common gaugino mass at the GUT scale,
the gaugino masses scale as
\beq
\alpha_3:\alpha_2:\alpha_1
\eeq
just 
as in gauge mediation. Again the bino is the LSP.
The gravitino mass is
of order the superpartner masses in this case.

These are a few possibilities for mediating supersymmetry breaking 
but by no means an exhaustive list.

 
\section{The MSSM spectrum}

\subsection{EWSB and the Higgs mass}
 
In the MSSM we have two Higgses, $H_U$ and $H_D$, which can get VEVs,
\beqa
\langle H_U\rangle =
\begin{pmatrix}
v_U
\\0
\end{pmatrix}~~~~
\langle H_D\rangle =
\begin{pmatrix}
0\\
v_D
\end{pmatrix}
\eeqa
Let's start in the supersymmetry limit (with no mu term).
The $D$ term must vanish, so the VEVs must be equal,
\beq
D=0~~~ \to~~~~
v_U=v_D
\eeq
The two Higgs fields contain 
8 real scalars. Of these, 
3 are eaten by $W^\pm$, $Z$.

Consider the heavy $Z$ supermultiplet. It contains
a heavy gauge boson which has 3 physical polarizations, and therefore
3 bosonic dof's and a Dirac fermion (4 dof's).
Therefore, in order to have the same number of fermion and boson dof's
there must be one more real scalar. This scalar comes from the Higgs fields.
The same holds for the $W^\pm$.
Thus, 3 real scalars ``join'' the heavy $W^\pm$, $Z$ supermultiplets.
In the limit of unbroken supersymmetry which we are assuming now,
all of these fields have masses $M_W$ or $M_Z$.

Thus, of the 8 real scalars in $H_U$ and $H_D$, 
2 neutral fields remain. One is the SM physical Higgs, $h$.
The other must be there because we have supersymmetry, and $h$ must reside
in a chiral supermultiplet. As we saw above, this multiplet contains
a complex scalar field.

Note that so far there is no potential for $h$. 
This is not surprising.
We haven't added any Higgs superpotential so the Higgs could only have a  
quartic 
from $V_D$.
But along the $D$-flat direction, the physical Higgs is massless.
Thus its mass must come from supersymmetry breaking !

Fortunately supersymmetry is broken---we have soft terms.
The Higgs potential comes from the following sources:
\begin{itemize}
\item The mu term: $W=\mu H_U H_D$,
\beq
\delta V=\vert \mu\vert^2 \vert  H_U\vert^2 +\vert \mu\vert^2 \vert  H_D\vert^2
\eeq
\item
The Higgs soft masses: 
\beq
\delta V=\tilde m^2_{H_U} \vert  H_U\vert^2 +
\tilde m^2_{H_D} \vert  H_D\vert^2
\eeq
so we need $m^2_{H_U}<0$ and/or $m^2_{H_U}<0$ 
\item
The $B\mu$ term:
\beq
\delta V= B\mu H_U H_D + {\rm hc}
\eeq
These are all quadratic terms.
\item Then we have quartic terms:
\beq
\delta V=\frac12 g_2^2D^I D^I + \frac12 g_1^2D_Y D_Y
\eeq
where

\beq
D^I =  H_U^\dagger \tau^I H_U - H_D^\dagger \tau^{I*} H_D 
\eeq
 and
\beq
D_Y = \sum_i Y_i \tilde f_i^\dagger  \tilde f_i 
+\frac12 (H_U^\dagger H_u - H_D^\dagger H_D)
\eeq
\end{itemize}

Recall we had two parameters, the  
two Higgs VEVs. We can  
trade them for:

\begin{enumerate}
\item
$\sqrt{v_U^2+v_D^2}$: determined by $W$ mass to be 246~GeV
\item
$
\tan\beta\equiv {v_U}/{v_D}
$
\end{enumerate}

Requiring a minimum of the potential determines:
\beqa
B\mu &=& \frac12 (m_{H_U}^2 + m_{H_D}^2 + 2\mu^2)\, \sin2\beta\\
\mu^2 &=& \frac{m_{H_D}^2 - m_{H_U}^2 \tan^2\beta}{\tan^2\beta -1} 
-\frac{M_Z^2}2
\eeqa
Thus,
for given $m_{H_U}^2$, $m_{H_D}^2$: $B\mu$ and $\mu$ are determined, and we have
two free parameters,
$\tan\beta$ and sign$(\mu)$.

Expanding around the VEVs we find that
the various scalars from $H_U$ and $H_D$ have the following masses (squared),
\beqa
H^\pm &:&  
M_W^2 + M_A^2 ~~~   ~~~  ~~~   ({\rm SUSY:} M_W^2)\nonumber\\
H^0 &:& 
\frac12\,(M_Z^2+ M_A^2) 
+\frac12\sqrt{(M_Z^2+ M_A^2)^2-4m_A^2M_Z^2\cos^22\beta} \nonumber\\
&~&~~~~~~~~~~~~~~~~~~~~~~~~~~~~~~~~~~~~({\rm SUSY:} M_Z^2)\nonumber\\
A^0 &:&  M_A^2= B\mu (\cot\beta+\tan\beta) ~~~~~({\rm SUSY}: 0)
\eeqa
and
for the physical Higgs,
\beq\label{htree}
m_h^2=\frac12\,(M_Z^2+ M_A^2) 
-\frac12\sqrt{(M_Z^2+ M_A^2)^2-4m_A^2M_Z^2\cos^22\beta}
\eeq
This is a  
{\bf PREDICTION:}
\beq
m_h \leq m_Z \vert \cos2\beta\vert\leq M_Z
\eeq
{\bf The measurement of the Higgs mass provides the first quantitative test
of the Minimal Supersymmetric Standard Model}. 
It fails.
However,
the result~\eqref{htree} is a tree-level result.
There are large radiative corrections to this result, mainly from stop loops.
In the decoupling limit
\beq
m_h^2 \sim m_Z^2\cos^22\beta +\frac{3m_t^2}{4\pi^2v^2}\, \left[
\log\frac{M_S^2}{m_t^2} + \frac{X_t^2}{M_S^2} 
\right]
\eeq
where
\beqa
X_t&=& A_t- \mu\cot\beta~~~~{\rm the ~LR ~stop ~mixing}\nonumber\\
M_S&=& \sqrt{m_{{\tilde t}_1}m_{{\tilde t}_2}} ~~~~{\rm the ~average ~stop 
~mass}\nonumber
\eeqa
This
can raise Higgs mass to around 130--150~GeV.
Thus for a 126~GeV Higgs we need
heavy stops and/or large stop A terms.
This is not very attractive. We wanted supersymmetry to solve the fine-tuning
problem, for which we need light stops. So the large Higgs mass typically 
implies some
fine-tuning.

In specific predictive models, like minimal GMSB, in which the stop mixing 
is small
(because there no A-terms at messenger scale), one needs
stops around 8~TeV, and because the other squarks and gluino masses are close 
by, all the colored superpartners are hopelessly heavy.
Thus, the Higgs mass sets a much stronger constraint on this framework 
than direct supersymmetry searches.

There's another important caveat. So far we did not add in any Higgs potential
on top of what the MSSM ``gave us''.
Let's compare this to the SM. 
In the
SM, we  add (by hand) a quartic Higgs potential, with a quartic 
coupling $\lambda$, to get the Higgs mass. 
Here we didn't have to: D-terms give a quartic potential.
As a result, there is no new parameter: $\lambda=g$.
We
could add a quartic interaction a la the SM.
To do that, we
must add at least one new field,
a SM singlet $S$,
with
\beq
W=\lambda S H_U H_D ~~\to ~~V=\lambda^2 (\vert H_U\vert^2 \vert H_D\vert^2 +\ldots)  
\eeq
This is called the
 Next to Minimal SSM (NMSSM).

We can pursue the
comparison to the SM at a deeper level.
In the SM, we put in  EWSB by hand. We had to put in a negative mass-squared
for the Higgs.
In the MSSM, EWSB can have  a dynamical origin.
Recall that we needed $\tilde m^2_{H_U}<0$ or $\tilde m^2_{H_D}<0$.
This happens (almost) automatically in supersymmetric theories, since
the RGEs drive the Higgs mass-squared negative!
The crucial contribution is due to 
the large Yukawa coupling of the Higgs to stops.

Now let's see why this happens.
Suppose we 
start with 
$\tilde m^2_{H_U}>0$
at the supersymmetry breaking scale.
The running gives
\beq
\frac{d}{dt} m^2_{H_U} \sim {\bf -} \frac{g^2}{16\pi^2}\, m^2_{1/2} 
{\bf +}  \frac{y_t^2}{16\pi^2}\, \tilde{m}^2_t  \,.
\eeq
The negative Yukawa contribution wins because  
\begin{enumerate}
\item The top Yukawa 
is large compared to the SU(2), U(1) gauge couplings.
\item
The stop is colored, so the Yukawa contribution is enhanced by a color 
factor (=3).
\end{enumerate}

Note that there are many scalars in the MSSM, so you could worry about their
masses-squared driven negative by the RGE.
However, the Higgs is special: it's an SU(3) singlet, so there is 
no large positive contribution from the gluino.
Furthermore, it has an order-1 Yukawa to the colored stop.
Thus only
the Higgs develops a VEV.

Let's summarize our results so far.
Putting aside the unpleasant 126~GeV Higgs mass
(which can be accounted for),
supersymmetry gives a very beautiful picture.
The MSSM (SSM + soft terms) has only log divergences:
the quadratic divergence in the Higgs mass-squared is cancelled 
by superpartners at $\tilde m$.
The tuning is then $\sim{M_Z^2}/{\tilde m}^2$ and
the hierarchy between the EWSB scale and the Planck/GUT scale
is {\bf stabilized}.

Furthermore:
starting with $\tilde m^2_{H_U}>0$ in the UV,
the running (from stops) drives it negative,
and electroweak symmetry is broken, with a scale proportional to $\tilde m$.

Finally, we remark that
with a SB sector that breaks supersymmetry dynamically,
the supersymmetry breaking scale is exponentially suppressed.
$\tilde m$ can naturally be around the TeV.
In this case,
the correct hierarchy between the EWSB scale and the Planck/GUT scale
is not only stabilized, but actually {\bf generated}.

Returning to the Higgs mass, 
with $m_h=126$~GeV
the {\bf Minimal} SSM is stretched: we often need heavy stops which
implies some level of tuning.
More practically,  discovery becomes more of a challenge.

Now that we understand supersymmetry breaking and EWSB
let's turn to the superpartner spectrum.

 \subsubsection{Neutralinos and charginos}
We have 4 neutral 2-component spinors: two gauginos and two Higgsinos
\beq
\tilde b\,,~~\tilde W^0\,,~~\tilde H_D^0  \,,~~\tilde H_U^0  
\eeq
with the mass matrix
\beqa
\begin{pmatrix}
M_1& 0 &-g_1 v_D/\sqrt2 & g_1 v_U/\sqrt2 \\
0&M_2& g_2 v_D/\sqrt2 & -g_2 v_U/\sqrt2 \\
 -g_1 v_D/\sqrt2 & g_2 v_D/\sqrt2 & 0 &\mu \\
 g_1 v_U/\sqrt2 & -g_2 v_U/\sqrt2  &\mu &0\\
\end{pmatrix}
\eeqa
Diagonalizing this matrix we find 
4 neutralino mass eigenstates: $\tilde\chi^0$~$i=1,\ldots,4$.
 
Similarly, there are  two charginos mass eigenstates which are combinations
of the charged Higgsino and wino, $\tilde \chi_i^\pm$~$i=1,2$.


\subsubsection{Sfermion spectrum}
Consider for example the up squarks.
There are
6 complex scalars: $\tilde{u}_{L i}$ and $\tilde{u}_{R a}$ 
with $i,a=1,2,3$ labeling the three generations.
The mass (squared) matrix is therefore a 6$\times$6 matrix:
\beqa
\begin{pmatrix}
m^2_{LL} & m^2_{LR}\\
m^{2\dagger}_{LR}& m^2_{RR}\\
\end{pmatrix}
\eeqa
where each of the blocks is 3$\times3$.

Consider $m^2_{U, LL}$. It gets contributions from:
\begin{enumerate}
\item the SSM Yukawa (supersymmetric)
\item the SUSY breaking mass-squared
\item the D-term (because $D\sim v_U^2-v_D^2 + \tilde q^\dagger T q +\cdots$)
\end{enumerate}
Thus,
\beq
m^2_{U,LL}= m_u^\dagger m_u + \tilde m^2_q + D_U 1_{3\times3}\,,
\eeq
and similarly for $m^2_{U, RR}$.

$m^2_{LR}$ gets contributions from:
\begin{enumerate}
\item the A term (supersymmetry breaking)
\item the $\mu$ term: \\
\beq
\left\vert
\frac{\partial W}{\partial H_D}
\right\vert^2 ~~\to~~
\frac{\partial W}{\partial H_D} = \mu H_U + y_U q u^c
\eeq
\end{enumerate}
so
\beq
m^2_{U,LR}= v_U\,(A_U^* -y_U \mu\cot\beta).
\eeq

The remaining sfermions (down-squarks, sleptons, sneutrinos) have a similar
structure.

 \subsubsection{Flavor structure}
Now let's consider the sfermion flavor structure,
starting with the up squarks as before.
Work in the quark mass basis (up, charm, top):
the Lagrangian contains the following:
\begin{itemize}
\item gaugino---$u_{L i}$---$\tilde u_{L j}$ couplings. Here the gaugino can be
either a gluino, a wino or a bino.
In our original Lagrangian, these are proportional to $\delta{ij}$.
We therefore say that the Lagrangian is given in the interaction basis.
Note that since we are in the fermion mass basis, this defines 
the L up squark, charm squark and top squark (stop).
For example, the L stop is the state that couples to a gluino and the 
doublet-top quark. 
\item gaugino---$u_{R a}$---$\tilde u_{R b}$ couplings.
Again, in our original Lagrangian, these are proportional to $\delta{ab}$.
\item \ldots
\item The up squark $6\times6$ mass matrix. 
This can in principle have an arbitrary structure.
In particular, the various $3\times3$ blocks need not be diagonal.
\end{itemize}
Diagonalizing the squark mass matrix, we get 6 mass eigenstates,
$\tilde u_I$, with $I=1,\dots,6$.
However, the gaugino---quark---squark couplings are no longer diagonal.
Writing these in terms of the up-squark mass eigenstates
we have in general
\beq
K_{iI} \tilde{g} u_{L i}-\tilde u_I\,,
\eeq
These mix  the different generations,
and $K_{iI}$ are the quark-squark mixing parameters.
Each squark (mass state) is a composition of the different flavor states.

Are sfermions degenerate? Is $m^2_{U,LL}\propto 1$?
That depends on the mediation of supersymmetry.
However, we don't understand the structure of fermion masses.
In fact this structure is very strange, suggesting a fundamental
theory of flavor.
If there is such a theory, it will also control the structure of  $m^2_{U,LL}$
and the other sfermion mass-matrices.

 \subsubsection{R-parity violating couplings}
So far, we merely generalized the SM gauge and Yukawa interactions.
In the SM, the Yukawa couplings were the only renormalizable couplings allowed.
Now however, there are  {\sl new} renormalizable Yukawa-like interactions
\beq
W= \lambda_{ijk} L_i L_j e^c_k + \lambda^\prime_{ijk} L_i Q_j d^c_k + 
 \lambda^{\prime\prime}_{ijk} u^c_i d^c_j d^c_k \,.
\eeq
These are the only terms we can add, nothing else is gauge invariant.
These terms are problematic. 
The first two terms break lepton number, the third breaks baryon number.
If they are all there we would get proton decay!
Note also that 
all these new terms break R-parity.
If we {\sl impose} R-parity, these dangerous terms are forbidden, and proton
decay can only arise from higher-dimension operators, much like in the SM.
But this may be overly restrictive. Certain flavor patters of R-parity
breaking operators are viable.

\section{LHC searches: general considerations}\label{lhc}
As we saw, supersymmetry is not a single model.
In the minimal supersymmetric extension of the standard model, the MSSM,
in which we add no new fields to the SM apart from those required 
by supersymmetry,
all the interactions are dictated by the SM and supersymmetry
(with the exception of R-parity violating couplings),
and the only freedom is in the soft supersymmetry breaking terms, 
which are  determined by the mechanism for mediating 
supersymmetry breaking.
Still, this allows for a wide variety of new and distinct signatures.
Therefore, while searches for specific models are useful, it's
also important to adopt a signature-based approach.

Here we will outline the main considerations to keep in mind.
\begin{enumerate}
\item Interactions:
There are only two sources of model dependence here.
\begin{itemize}
\item 
R-parity violating (RPV) couplings:
If R-parity is conserved, superpartners are produced in pairs, and
each superpartner decays to a lighter superpartner plus SM particles.
Thus any decay chain ends with the stable LSP.
In the presence of RPV couplings, a single superpartner
can be produced.
There are strong bounds on RPV couplings.
Still, single superpartner production via a small RPV coupling
may be competitive because of the kinematics.
Most superpartners   decay through the usual R-parity conserving couplings
(gauge or Yukawa), with the exception of the LSP, which can only decay through
RPV couplings.
\item Squark and/or slepton flavor mixing:
For general sfermion mass matrices, the 
gaugino-sfermion-fermion couplings may mix different flavors.
This could affect both production and decay. 
\end{itemize}
\item Superpartner masses:
\begin{itemize}
\item The hierarchy between colored and non-colored superpartners:
In most mediation schemes, colored superpartners are heavier (roughly by factors
of a few to 10).
Unless there is a huge hierarchy between colored and non-colored
superpartners, the production of squarks and gluinos dominates
at the LHC. As the hierarchy increases, the production of EWK gauginos
(ewkinos) and or sleptons becomes more competitive.
\item Flavor structure: superpartners of the same gauge charges may have
generation dependent-masses. Thus for example, the L-handed up squark 
can have a different mass from the L-handed charm or top squarks.
An arbitrary flavor structure leads to flavor changing processes which,
especially when the 1st and 2nd generations  are involved, are
stringently constrained. However, roughly speaking, the constrained quantities
involve the products of the mass splittings and the flavor mixings $K_{iI}$,
so models with some degree of mass degeneracy and some alignment of fermion
and sfermion mass matrices are allowed.
These would affect both the production of sfermions and their decay.  
\end{itemize}
\end{enumerate}
The (N)LSP plays a special role in determining the collider signatures 
of supersymmetry. Both the mass spectrum and its interactions are relevant here.
\begin{itemize}
\item The (N)LSP lifetime: 
The LSP is stable if R-parity is conserved. 
In the presence of 
RPV couplings, the LSP can decay to SM particles, but its lifetime may be long
because the sizes of these couplings are constrained.
Finally, the LSP may have only very weak couplings to the SM, as is the case
of the gravitino. Superpartners produced at the LHC will decay to the lightest
superpartner charged under the SM, which is called the NLSP (next to lightest
superpartner), which in turn decays to the LSP. Clearly, since the latter
is only weakly coupled to the SM, the NLSP can be long lived.
Different models span the whole range from propmt NLSP decays to NLSPs
which are long-lived on detector scales.
\item The (N)LSP charge. The (N)LSP can be either neutral 
(eg a neutralino or sneutrino), 
charged (eg a slepton), 
or even colored (eg the gluino).
Naturally, the precise identity of the (N)LSP plays an key role in determining
the signatures of supersymmetry at the LHC. The signatures of a spectrum
with a pure bino (N)LSP are very different from the   signatures of a spectrum
with a pure Higgsino (N)LSP even though both are neutral.
\end{itemize}
If the (N)LSP is neutral and long lived, superpartner production is accompanied
by missing energy. If it's charged and long lived, it behaves like a heavy muon,
and $dE/dx$ and time-of-flight measurements must be used to distinguish it
from a muon. If it's colored and long lived, it will hadronize in the
detector, and subsequently either stop in the detector or traverse 
the entire detector. If it decays inside the detector, its charge and
lifetime determine the specific signature, ranging from
a disappearing track
to a displaced vertex, with or without missing energy.
Thus, supersymmetry has motivated a wide variety of ingenious approaches 
for searching
for new physics.
Whether supersymmetry is there or not, this net of searches will hopefully
lead to new discoveries!

\section*{Acknowledgements}
I thank the organizers of ESHEP2014 for putting together a very pleasant and
stimulating school.


\end{document}